\documentclass[prl,aps,twocolumn,superscriptaddress]{revtex4-1}
\usepackage{latexsym}
\usepackage{graphicx}
\usepackage{verbatim}
\usepackage{multirow}
\usepackage{amsmath}
\usepackage{mathrsfs}
\usepackage{float}
\usepackage{feynmf}

\newcommand{\be}{\begin{equation}}
\newcommand{\ee}{\end{equation}}
\newcommand{\ba}{\begin{eqnarray}}
\newcommand{\ea}{\end{eqnarray}}
\newcommand{\baa}{\begin{eqnarray*}}
\newcommand{\eaa}{\end{eqnarray*}}

\def\be{\begin{equation}}
\def\ee{\end{equation}}
\def\bea{\begin{eqnarray}}
\def\eea{\end{eqnarray}}

\def\C60{A$_x$C$_{60}$}

\def\HgCu3{HgCa$_2$Cu$_3$O$_{8+y}$}
\def\HgCu4{HgBa$_2$Ca$_3$Cu$_4$O$_{10+y}$}
\def\TlCu{Tl$_2$Ba$_2$CuO$_{6+\delta}$}
\def\TlCu3{Tl$_2$Ba$_2$Ca$_2$Cu$_3$O$_{10+y}$}
\def\TlCu4{Tl$_2$Ba$_2$Ca$_3$Cu$_4$O$_{12+y}$}

\def\BiCu3{Bi$_2$Sr$_2$Ca$_{2}$Cu$_3$O$_y$}

\def\8LSCO{La$_{1.88}$Sr$_{.12}$CuO$_4$}
\def\110LNSCO{La$_{1.5}$Nd$_{0.4}$Sr$_{0.1}$CuO$_{4}$}
\def\stage4LCO{La$_{2}$CuO$_{4+\delta}$}
\def\Y248{YBa$_2$Cu$_4$O$_8$}

\def\NbSe2{NbSe$_2$}
\def\TaSe2{TaSe$_2$}
\def\TiSe2{TiSe$_2$}

\begin{document}
\title{Study of non-Fermi Liquid behavior from partial nesting in multi-orbital superconductors.}
 \author{Chandan Setty}
\affiliation{Department of Physics and Institute of Condensed Matter Theory, University of Illinois 1110 W. Green Street, Urbana, IL 61801, USA}
\author{Philip W. Phillips}
\affiliation{Department of Physics and Institute of Condensed Matter Theory, University of Illinois 1110 W. Green Street, Urbana, IL 61801, USA}
\begin{abstract}
Partial nesting between two connected or disconnected regions of the Fermi surface leads to fractional powers of the Coulomb scattering lifetime as a function of temperature and frequency. This result is first demonstrated for a toy band structure where partial nesting occurs within a single band and between different regions of the Brillouin zone. A comparison is then made to a multiband scenario by studying the scattering rate of an effective two orbital model that was proposed in the context of multi-orbital superconductors. In the process, various \textit{model independent} factors affecting the temperature exponent, $n$, are identified. The logarithmically divergent contributions of the lowest order vertex correction to the multi-orbital susceptibility, and the role played by nesting in suppressing these divergences is analysed. The relevance of these results is discussed keeping the recently observed anomalous resistivity in the $Co$ doped Iron superconductor $LiFeAs$ as a backdrop.
\end{abstract}
\maketitle
\textit{Introduction:} While deviations from the standard theory of metals arise typically from strong local correlations, they also come into play anytime two pieces of the Fermi surface coincide when shifted by a wave vector $\vec Q$.  This phenomenon of nesting, well documented in magnetism,  defined by $\epsilon_{\vec k} = - \epsilon_{\vec k + \vec Q}$, is relevant for many classes of high $T_c$ materials including the Cuprates\cite{Spicer1993} and the Iron superconductors\cite{Takahashi2009}. Given that most of the proposals so far for the pairing mechanism in both these classes of high $T_c$ materials involve the distribution of magnetic fluctuations in the Brillouin zone, and that the topology of the Fermi surface plays a key role in controlling the magnitude of the spin susceptibility, a nested Fermi surface can give qualitatively different results when compared to an unnested one. This was first demonstrated by Virosztek and Ruvalds \cite{Ruvalds1990,Ruvalds1999} where they showed that the inverse Coulomb scattering life time in the presence of a perfectly nested Fermi surface behaves linearly as a function of temperature. These results were extended by Schlotmann\cite{Schlottmann2003,Schlottmann2006,Schlottmann2007} to other parallel cases away from nesting .     \\
\newline
Recently Dai et.al \cite{Ding2015} performed a series of experiments including ARPES, optical spectroscopy, NMR and transport on a variety of $Co$ doped concentrations of $LiFe_{1-x}Co_xAs$. After studying the band structures, Fermi surfaces, spin lattice relaxation rates and the Coulomb scattering life times for all the $Co$ concentrations, they\cite{Ding2015} concluded that there was a Fermi liquid to Non Fermi liquid to Fermi Liquid transition as a function of the $Co$ concentration. In the non-Fermi liquid regime, fractional powers of the self energy vs temperature were measured. These observations occur simultaneously as  the hole pockets at the $\Gamma$ point start from being larger than the electron pockets at the $M$ point in the folded Brillouin zone, then become smaller with electron doping while going through a transition where they are close to being perfectly nested. However, the $T_c$ of the material monotonically reduces with $Co$ doping indicating no correlation with the magnitude of spin fluctuations. As a result, these experiments point to nesting playing a crucial role in driving the Fermi liquid to non-Fermi liquid transition.  \\

\begin{figure}[h!]
\includegraphics[width=0.5\textwidth]{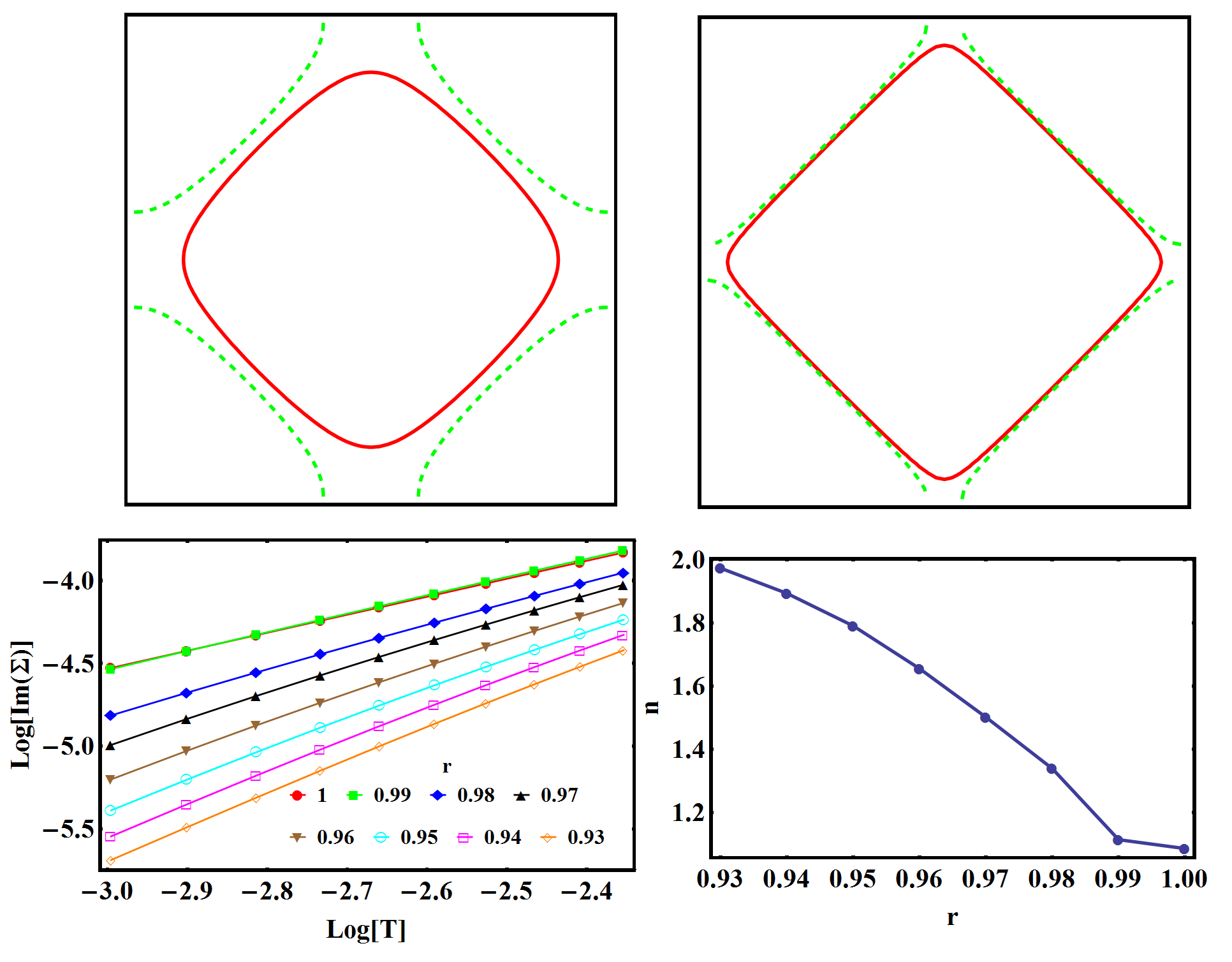}
\caption{Temperature dependence of the imaginary part of the Coulomb self energy with the nesting parameter ($r$) for a one-orbital toy model: (top-left) Fermi surface (solid, red) far away from nesting ($r=0.9$). The contour shifted by the nesting wave vector $(\pi,\pi)$ is shown in dashed green. (top-right) Fermi surface close to the perfect nesting condition ($r=1$). (bottom-left) Log-log plot of the imaginary part of the Coulomb self energy as a function of temperature for different values of the nesting parameter $r$. (bottom-right) Exponent ($n$) of the temperature extracted from the previous plot as a function of $r$.  }\label{OneOrbital}
\end{figure}

Although other factors are undoubtedly present in these materials, we explore in this paper the role nesting can play in driving a Fermi liquid to non-Fermi liquid transition in an effective two-band model of Iron superconductors.   In the perturbative regime, we calculate the electron self energy and the electron scattering lifetime as a function of temperature.  In general the temperature dependence is of the power-law kind, $T^n$.  Our findings for the single orbital toy model show that the exponent $n$ decreases continuously from the Fermi Liquid value of two to unity as the nesting condition improves. For the effective two orbital model with a nesting wave vector connecting two disconnected hole and electron pockets, the exponent $n$ goes through a minimum for a certain doping value consistent with perfectly nested pockets. In an attempt to understand the minimum value of the exponent reached, we identify various \textit{model independent} factors which affect the temperature exponent, $n$, in a multi-orbital system. In particular, we find that the presence of intra-band scattering, matrix element effects and the invalidity of replacing the fully momentum dependent susceptibility with that evaluated at the nesting wave vector alone (nesting approximation\cite{Ruvalds1990,Ruvalds1999}), pushes the value of $n$ to be larger than unity and  above the experimentally observed value of 1.35.  By studying the lowest order vertex corrections to the bare susceptibility for the multi-orbital scenario, we numerically find a logarithmic divergence at low temperatures which is systematically suppressed as we move away from the perfectly nested condition. \\
\newline
The following paragraphs are organized as follows: we start by writing out the expressions for the self energy in a generic multi-orbital Hamiltonian. We then present our results for the toy model and those for the effective two oribital model proposed in the context of the Iron superconductors. This is followed by our analysis of the various factors determining the exponent, $n$. Finally, we derive expressions for the lowest order multi-band vertex corrections and end with our conclusions.\\
\newline
\textit{Theory:} We begin by with a many body Hamiltonian with inter- and intra- orbital hoppings and electron-electron interactions given by
\begin{equation}
H = \sum_{ k\alpha\beta} \epsilon_{\alpha \beta}(\vec k) c_{k \alpha}^{\dagger}c_{k \beta} + \frac{1}{2} \sum_{ \substack{k  k' q\\ \alpha \beta \gamma \delta}}U_{\alpha \beta \gamma \delta} c_{ k+ q \alpha}^{\dagger} c_{ k'- q \beta}^{\dagger}  c_{ k' \gamma} c_{ k \delta}.
\end{equation}
Here $c_k, c_k^{\dagger}$ are the electron annihilation and creation operators, $\epsilon_{\alpha\beta}(\vec k)$ are the hopping matrix elements, $\alpha, \beta, \gamma, \delta$ are the orbital indices and $U_{\alpha\beta\gamma\delta}$ are the interaction matrix elements. In the analysis to follow, we will first consider a single orbital model for simplicity and then proceed to an effective two-orbital model to apply to the Iron-based superconductors. We will also use an orbital independent interaction given by $V_{\vec q}$  which is the bare Coulomb repulsion between the electrons without the $\vec q =0 $ component. The non-interacting Green function in the orbital basis is given  by
\begin{equation}
G_{\alpha\beta}^{(0)}(\vec k, i \omega_n) = \sum_{m} \frac{u_{m\alpha}^*(\vec k) u_{m\beta}(\vec k)}{i\omega_n - \mathscr{E}_m(\vec k)}.\label{Green}
\end{equation}
Here, $u_{m\alpha}$ are the matrix elements connecting the band and the orbital basis, $i\omega_n$ are the fermionic Matsubara frequencies, $\mathscr{E}_m$ are the band energies labelled by the band index $m$. With the knowledge of the non-interacting Green function at hand and using standard many -body perturbation approaches in the interaction parameter, we write the interacting Green function as
\begin{equation}
\hat{G}(\vec k, i k_n) = \hat{G}^{(0)}(\vec k, i k_n) + \hat{G}^{(0)}(\vec k, i k_n)\hat{\Sigma}(\vec k,  i k_n) \hat{G}(\vec k, i k_n),
\end{equation}
\begin{figure}[h!]
\includegraphics[width=0.45\textwidth]{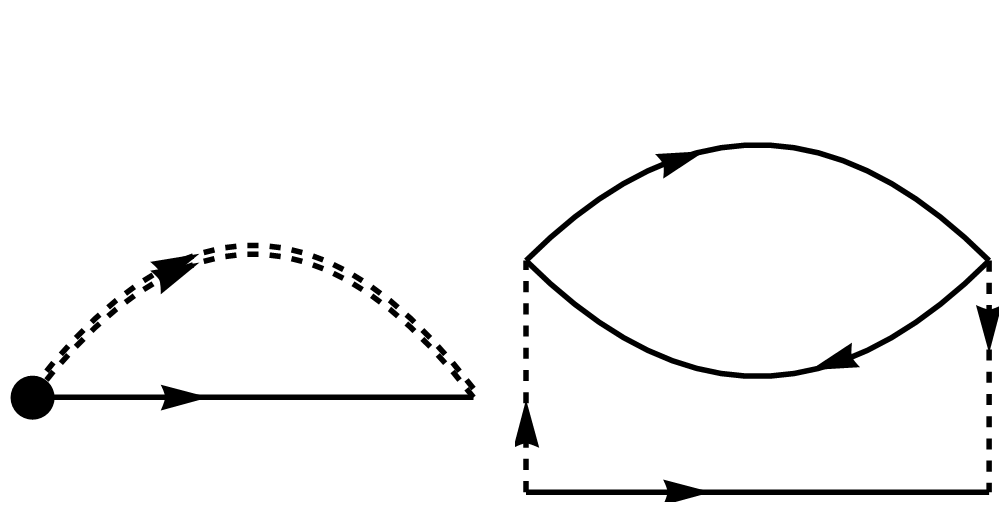}
\caption{Diagrams contributing to the Coulomb self energy. Left: full self energy contribution with a renormalized vertex (shaded blob) and interaction parts (double dashed line). Right: one of the contributing terms to the RPA self energy $\Sigma_{RPA}(\vec k, \omega)$ obtained by inserting a single fermion bubble into the interaction line along with a bare vertex. }\label{Feynman}
\end{figure}
which must be solved recursively to obtain $\hat{G}(\vec k, ik_n)$. The hat on top of the physical quantities signifies a matrix character with the matrix elements denoted by the orbital index. The above Dyson-like equation for $\hat{G}(\vec k, ik_n)$ defines the self energy $\hat{\Sigma}(\vec k, ik_n)$. The imaginary part of the total self energy (including the individual orbital contributions) under the Random Phase Approximation (RPA) reduces to
\begin{eqnarray}
 \Sigma''(\vec k, \omega)&=& \sum_{\substack{\alpha \beta\\ m \vec q}} V_{RPA}(\vec q)^2 \chi_o''\left(\vec q, \omega - \mathscr{E}_m(\vec k -\vec q)\right)\\ \nonumber
&& \times u_{m\alpha}^*(\vec k - \vec q) u_{m\beta}(\vec k -\vec q)\\ \nonumber
&&\times \left(n_f(-\mathscr{E}_m(\vec k -\vec q)) +n_b(\omega-\mathscr{E}_m(\vec k -\vec q))\right).
\end{eqnarray}
$V_{RPA}(\vec q)$ is the RPA renormalized Coulomb interaction, $n_f(x)$ and $n_b(x)$ are the Fermi and Bose distribution functions and $\chi_o(\vec q, \omega)$ is the total multi-orbital susceptibility:\begin{eqnarray}
\chi_o(\vec q, i\omega_n) &=& \sum_{\substack{a,b\\c,d}}\chi_{abcd}(\vec q, i \omega_n)\\
\chi_{abcd}(\vec q, i \omega_n) &=& \sum_{\vec p, i p_n} G_{ac}^{(0)}(\vec p, ip_n)G_{bd}^{(0)}(\vec p + \vec q, ip_n + i\omega_n).\nonumber
\end{eqnarray}
The imaginary part of the self energy gives us the electron-electron scattering lifetime, $\tau$, through the relation $\tau(\vec k, \omega) = -1/2 \Sigma''(\vec k, \omega)$. The derivation of the above RPA expressions for the self energy is represented diagramatically in Fig. \ref{Feynman}.
\newline
\newline
\textit{Results}: We have obtained results for two representative cases: (i) a single orbital model and (ii) an effective two-orbital model for the Iron based superconductors. For the one orbital model we choose a band structure described by $\epsilon(\vec k) = t_1(cos k_x + cos k_y) + (t_2- r)(cos k_x cos k_y) - (\mu - 2 r)$. We choose the parameters as $(t_1, t_2, \mu) = (1.0,1.0, 2.0)eV$. The value of $r$ is assumed to change with an external parameter such as doping. The Fermi surface of such a band structure has a nesting wavevector of $\vec Q = (\pi, \pi)$ for values of $r$ close to unity, i.e, $\epsilon(\vec k) = -\epsilon(\vec k+ \vec Q)$ when $r = 1$. As $r$ deviates from unity, the nesting slowly deteriorates. The Fermi surface close to and far away from nesting is shown in the top panels of Fig \ref{OneOrbital}.
 The bottom left panel in Fig \ref{OneOrbital} shows a log-log plot of the imaginary part of the self energy as a function of temperature and its simultaneous variation with the nesting parameter $r$. For all the plots we have chosen a value of $\vec k$ on the Fermi surface close to the $(\pi/2, \pi/2)$ point and an energy close to the Fermi level with $\omega = 2 meV$. Clearly, there is a strong dependence of the slope of the straight lines as a function of $r$. The slope of the straight line gives the exponent $n$ in the temperature variation of the imaginary part of the self energy i.e. $\Sigma''(\vec k, \omega) \propto T^n$. As a result, the exponent $n$ changes continuously from $n=2$ (Fermi liquid behavior)to $n=1$ (non-Fermi liquid behavior) as nesting becomes more and more perfect ($r \rightarrow 1$). This dramatic variation consisting of different intermediate values of $n$ is shown in the bottom right panel of Fig \ref{OneOrbital}. The case of perfecting nesting was first discussed by Virosztek and Ruvalds\cite{Ruvalds1990} who show that in the presence of nesting, the susceptibility bubble appearing in Fig \ref{Feynman} scales as a function of $\omega/T$, while in the case of a Fermi liquid, the susceptibility is linear in $\omega$ and  is essentially temperature independent. This qualitatively different behavior of the susceptibility in case of perfect nesting leads to an increased electron-electron interaction induced scattering and, quantitatively, obtains a linear temperature dependence of the self energy; in contrast, a close to quadratic temperature dependent behavior is seen in a two-dimensional Fermi liquid in a limit where $\hbar \omega<<k_B T<<E_f$. \\
\begin{figure}[h!]
\includegraphics[width=0.52\textwidth]{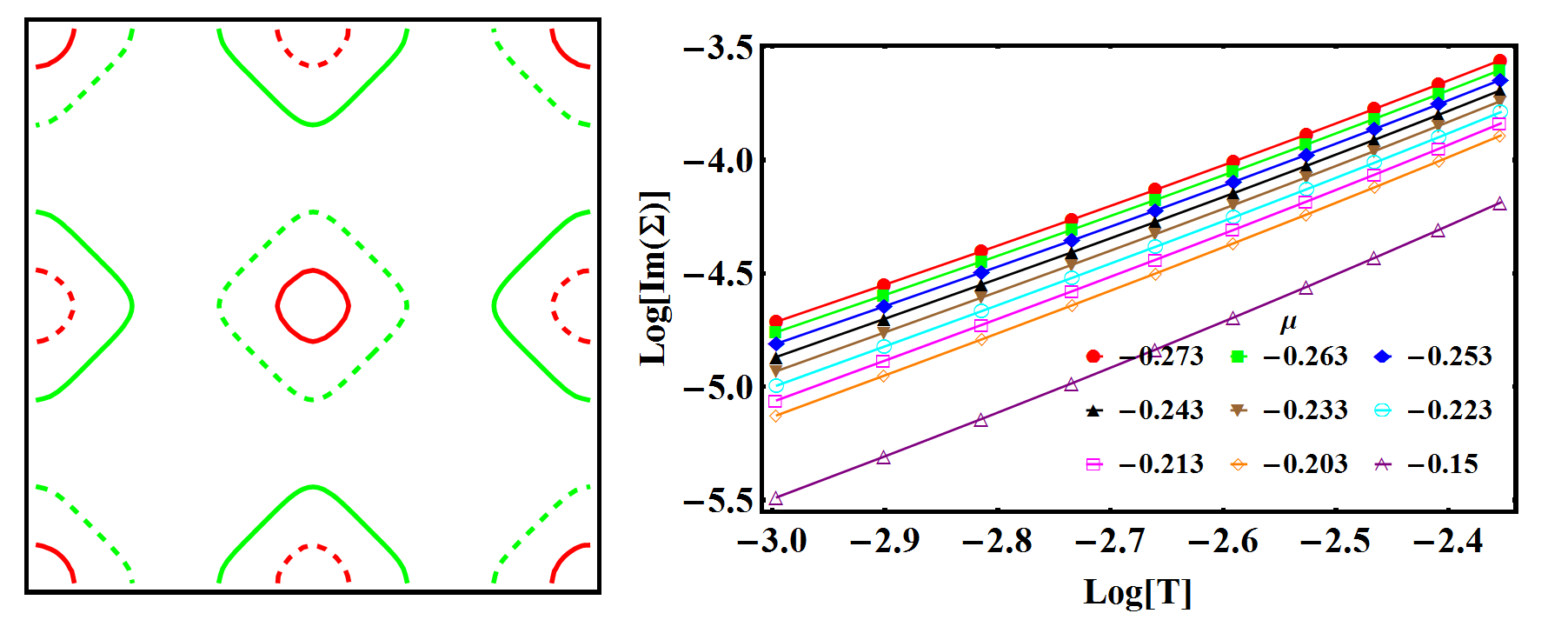}
\caption{(Left) Fermi surface of the model in ref\cite{Hu2012-S4} away from nesting when the hole pocket at the $\Gamma$ point(red, solid) has a smaller area than the electron pocket at the $X$ point (green,solid). The dashed lines are the Fermi surfaces shifted by the nesting vector $(\pi,0)$. (Right) Corresponding log-log plots of the imaginary part of the Coulomb self energy vs temperature for several dopings in this regime. $\mu = -0.273$ is close to the perfect nesting of the electron and hole pockets. }\label{FeSCNesting1}
\end{figure} 
\newline
We consider next an effective two-orbital model relevant to the Iron-based superconductors. We  make a quick recap of the $S_4$ symmetric model described by the authors in \cite{Hu2012-S4}. Our starting point is the following Hamiltonian similar to \cite{Zhang2008} on a single copy containing the $x'z$ on the $A$ sublattice and $y'z$ ($x'$ and $y'$ are along the diagonals to the $Fe-Fe$ bonds) on the $B$ sublattice coupled to each other through the $As$ atoms in between:
\begin{equation}
H_0 =  \sum_{\textbf{k},\sigma}\psi_{\textbf{k},\sigma}^{\dagger} \left(\epsilon_+(\textbf{k}) - \mu)1 + \epsilon_-(\textbf{k})\tau_3 + \epsilon_{xy}(\textbf{k}) \tau_1\right)\psi_{\textbf{k},\sigma} \label{BandHamiltonian}
\end{equation}
with $\tau_i$ the Pauli matrices and  $\psi_{\textbf{k},\sigma}^{\dagger} = (c_{1,k}^{\dagger}, c_{2,k}^{\dagger})$ with $c_{1,k}^{\dagger}$ and $c_{2,k}^{\dagger}$  the electron creation operators at the sublattice sites $A$ ($x'z$ orbital) and $B$ ($y'z$ orbital) respectively.  The band parameters are
\begin{eqnarray*}
\epsilon_{\pm}(\textbf{k}) &=& \frac{\epsilon_x(\textbf{k}) \pm \epsilon_y(\textbf{k})}{2}\\
\epsilon_x(\textbf{k}) &=& 4 t_s cos k_x cos k_y - 4 t_d sin k_x sin k_y \\
&&+ 2 t_{3s} ( cos 2 k_x + cos 2 k_y) \\
&&+ 2 t_{3d} (cos 2 k_x - cos 2 k_y)\\
\epsilon_y(\textbf{k}) &=& 4 t_s cos k_x cos k_y + 4 t_d sin k_x sin k_y \\
&&+  2 t_{3s} ( cos 2 k_x + cos 2 k_y) \\
&&+ 2 t_{3d} (cos 2 k_x - cos 2 k_y) \\
\epsilon_{xy}(\textbf{k})&=& 2 t_1 (cos k_x + cos k_y)
\end{eqnarray*}
with  $t_1=0.24, t_2 =0.52, t_2' = -0.1, t_{s,d} = (t_2 \pm t_2')/2, t_{3s} = t_{3d} \sim 0$, and $\mu = -0.273$ for perfect nesting. The matrix elements and electron operators ($d_{1,k}$ and $d_{2,k}$) for the other copy with the $y'z$ orbital on the $A$ and $x'z$ on the $B$ sublattice can be obtained by performing the $S_4$ symmetry transformation as demonstrated in \cite{Hu2012-S4}.\\
\newline
\begin{figure}[h!]
\includegraphics[width=0.52\textwidth]{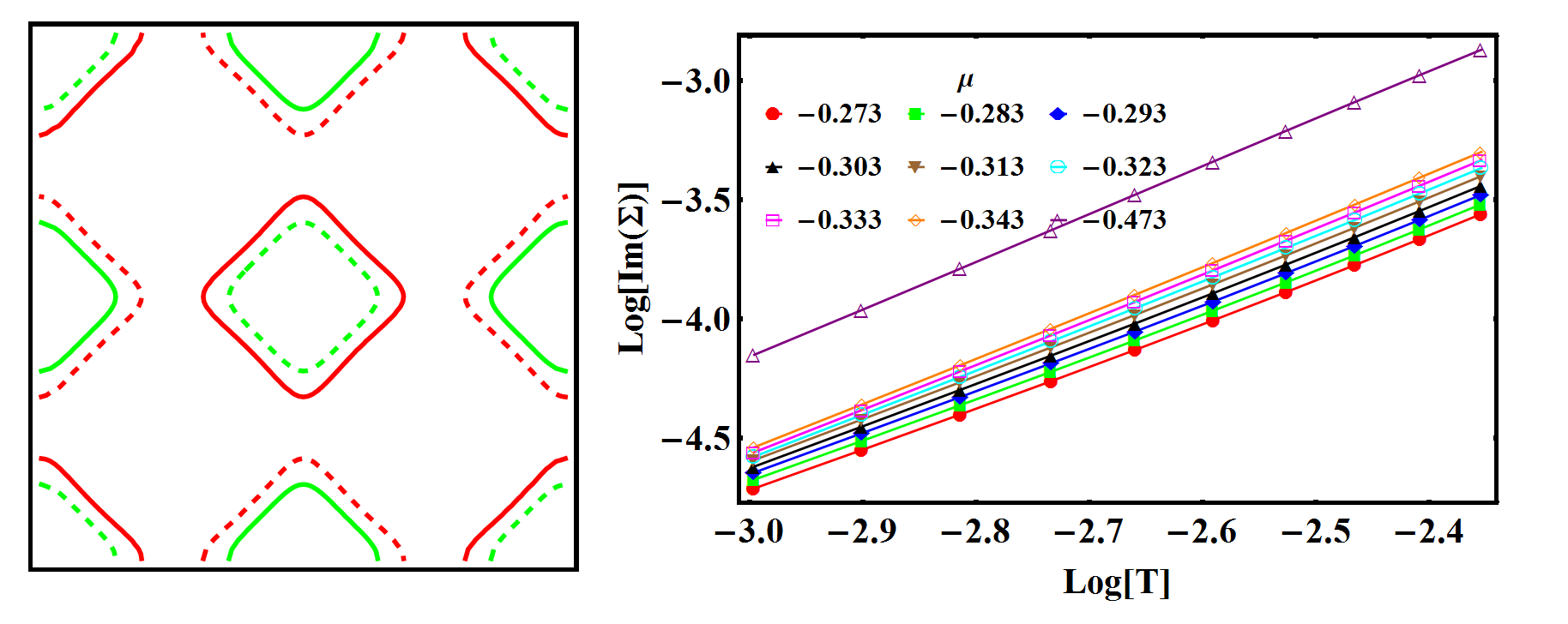}
\caption{(Left) Fermi surface of the model in ref\cite{Hu2012-S4} away from nesting when the hole pocket at the $\Gamma$ point(red, solid) has a larger area than the electron pocket at the $X$ point (green,solid). The dashed lines are the Fermi surfaces shifted by the nesting vector $(\pi,0)$. (Right) Corresponding log-log plots of the imaginary part of the Coulomb self energy vs temperature for several dopings in this regime. $\mu = -0.273$ is close to perfect nesting of the electron and hole pockets.}\label{FeSCNesting2}
\end{figure} 
Fig \ref{FeSCNesting1} (left) shows a contour plot of the Fermi surface for the model described above. The value of $\mu$ is chosen to be larger than -0.273 and is electron doped. At $\mu = -0.273$ the hole and electron pockets fall exactly on top of each other when the band structures are shifted by $\vec Q = (\pi, 0)$ and correspond to perfect nesting. When $\mu$ is increased to values above -0.273, the hole pocket has a smaller area than the electron pocket and results in a poor nesting condition between the hole and electron pockets. Fig \ref{FeSCNesting1} (right) shows a log-log plot of the imaginary part of the multiorbital self energy as a function of temperature. The different curves correspond to various values of $\mu$ from $-0.273$ to $-0.15$. The slope of the straight lines (the exponent $n$) gradually increases from $n=1.75$ to $n=2$ as we move from $\mu = -0.273$ to $\mu = -1.5$ signalling a decreased scattering rate when the nesting condition deteriorates. \\

\begin{figure}[h!]
\includegraphics[width=0.52\textwidth]{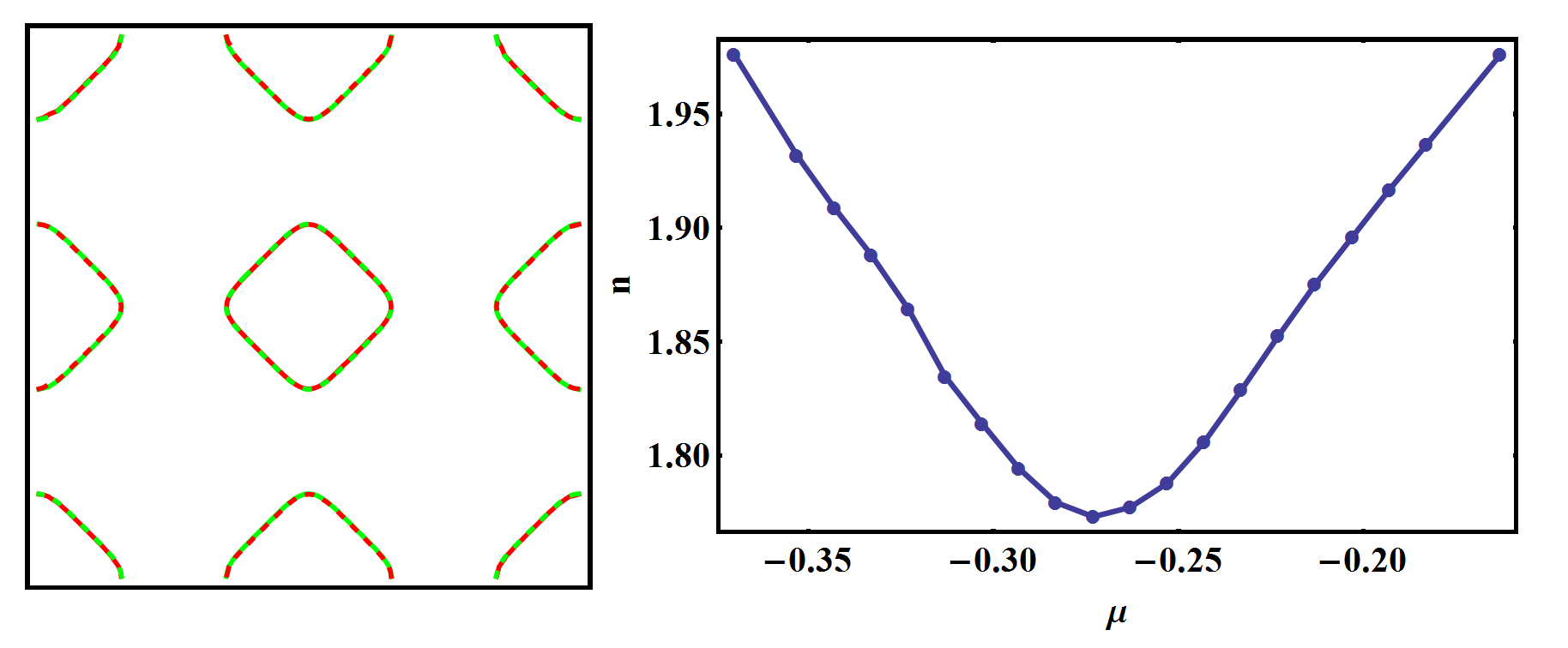}
\caption{ (Right) Exponents ($n$) of the temperature extracted from the log-log plots in figs 3 and 4 as a function of doping $\mu$. The minimum corresponds to close to perfect nesting of the electron and hole pockets (left).}\label{FeSCNesting3}
\end{figure}

Similar behavior is observed in the opposite case of hole doping. Fig \ref{FeSCNesting2} illustrates this scenario where the hole pockets become larger than the electron pockets (Fig \ref{FeSCNesting2} (left)). As the value of $\mu$ is decreased from -0.273 to -0.473, the nesting effect between the hole pocket and the electron pocket decreases resulting in a longer life time due to reduced electron-electron scattering. Thus an increased slope from $n=1.75$ to $n=2$ is obtained when $\mu$ is decreased to -0.473 (Fig \ref{FeSCNesting2}, Right). The behavior of the exponent $n$ is more clearly shown in Fig \ref{FeSCNesting3} (right). The value of $n$ starts out from being a Fermi liquid value of two when hole doped and then decreases in a continuous fashion to a mininum value (around $1.75$) when the hole pocket at the $\Gamma$ point and the electron pocket at the $X$ point are well nested. Finally it increases back to two when electron doped. The case of perfectly nested Fermi surface is shown in Fig \ref{FeSCNesting3} (left). It is important to note that, unlike the one-orbital scenario where the exponent reached a value of unity with perfect nesting, the exponent in this model decreases no further than 1.75. In the following paragraphs, we analyse different factors that come into play in setting this apparent lower bound.\\
\newline
It is deceptively natural, at this point, to attribute the larger than expected experimentally observed (and the above numerically evaluated) value of $n$ to details of the band structure. Iron superconductors are multiorbital systems with convoluted band structures and competing orders; therefore, it is possible to imagine that the large value of $n$ could stem from the fact that the hole pockets are many times flatter than the electron pockets near the Fermi surface, or that such a feature could result in improper nesting at the Fermi surface. Moreover, one could be led to believe that the volume of the Fermi surface being perfectly nested might play a role in determining the value of the exponent. However, none of these factors have any affect. To see this one can extend the same argument put forward originally for the one band scenario in ref \cite{Ruvalds1990}. The imaginary part of the bare multi-orbital non-interacting susceptibility (Lindhard function) is given by
\begin{eqnarray}
\chi_{0\alpha\beta}''(\vec q, \omega) &=& 2\pi\sum_{m m'}\int\frac{d^2 k}{(2 \pi)^2} \mathscr{L}_{mm'}^{\alpha\beta}(\vec k, \vec q) \nonumber \\
&&\times\left(n_f(\epsilon_m(\vec k)) - n_f(\epsilon_{m'}(\vec k + \vec q))\right)\nonumber \\
&&\times \delta^{(2)}\left(\omega + \epsilon_{m}(\vec k) - \epsilon_{m'}(\vec k + \vec q)\right), \label{Susceptibility}
\end{eqnarray}
where $\mathscr{L}_{mm'}^{\alpha\beta}(\vec k, \vec q)$ are the band matrix elements and $\epsilon_{m}(\vec k)$ are band energies, with $m,m'$ and $\alpha, \beta$ denoting band and orbital indices respectively. To study the effect of the difference in curvature between two perfectly nested bands at the nesting wave vector $\vec Q$, we can make the substitution $\epsilon_{m'}(\vec k + \vec Q) = - s \epsilon_{m}(\vec k)$, where $s$ is a scalar number. Such a substitution preserves  the perfect nesting condition at the Fermi surface but makes the curvature of the two bands dissimilar. Converting the momentum integrals in terms over energy and assuming a constant density of states and simplifiying using the delta functions, it can be numerically shown that such a substitution does not alter the liinear $T$ dependence of the Coulomb scattering rate.
\begin{figure}[h!]
\includegraphics[width=0.5\textwidth]{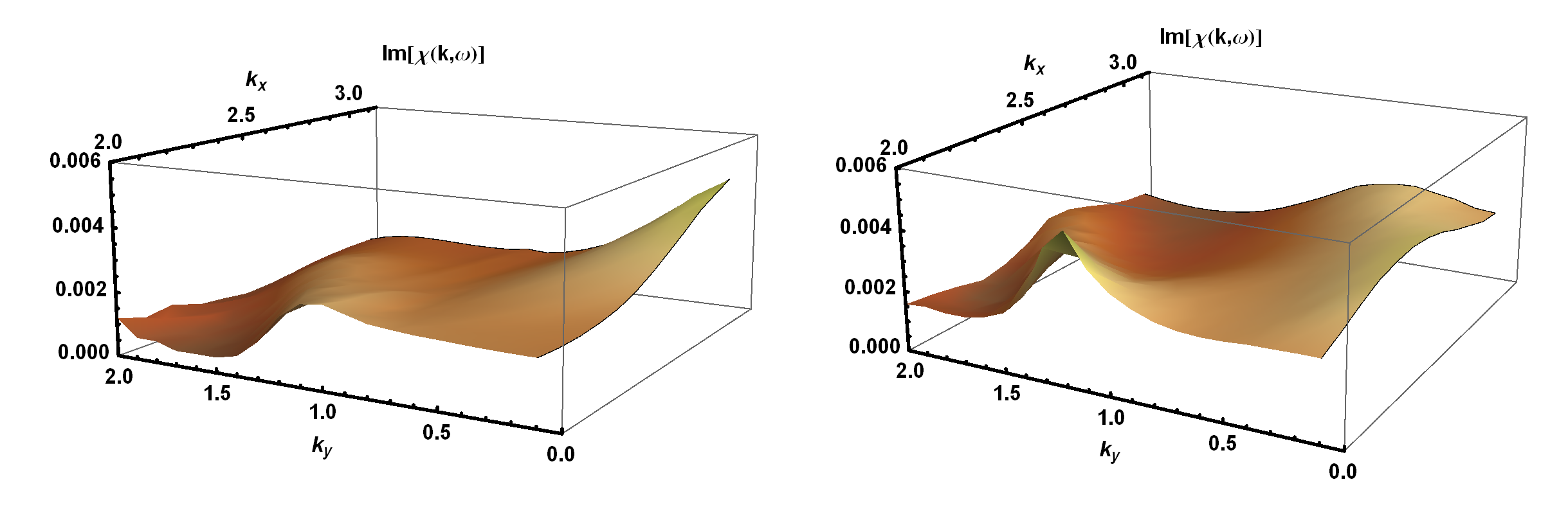}
\caption{Three dimensional plot of the imaginary part of the total susceptibility close to the nesting wave vector $(\pi,0)$ in the Brillouin Zone for the effective model. (Left) When the chemical potential $\mu = -0.273$ eV close to perfect nesting and (Right) when the chemical potential is far away from nesting. The response frequency $\omega$ and temperature $T$ are chosen as 2 meV and 50 meV respectively. }\label{Chi}
\end{figure}
Similarly, changing the area of the nested Fermi surface while maintaining the condition of perfect nesting does not affect the scattering rate. This is because the condition $\epsilon_m(\vec k) = - \epsilon_{m'}(\vec k + \vec Q)$ is satisfied regardless of the area of the Fermi surface nested. Finally, the ARPES measurements in ref\cite{Ding2015} (within the experiment's momentum and energy resolution) make it clear that at a $Co$ doping of $\approx 0.12$, the hole and electron pockets are close to perfect nesting ruling out any Fermi surface dissimilarities being responsible for pushing the exponent $n$ to be larger than unity. It is, therefore, natural to ask what factors really control the value of $n$ in our calculation. Here we point out that it is possible to extract model-independent features which affect the Coulomb scattering rate.

\begin{figure}[h!]
\includegraphics[width=0.5\textwidth]{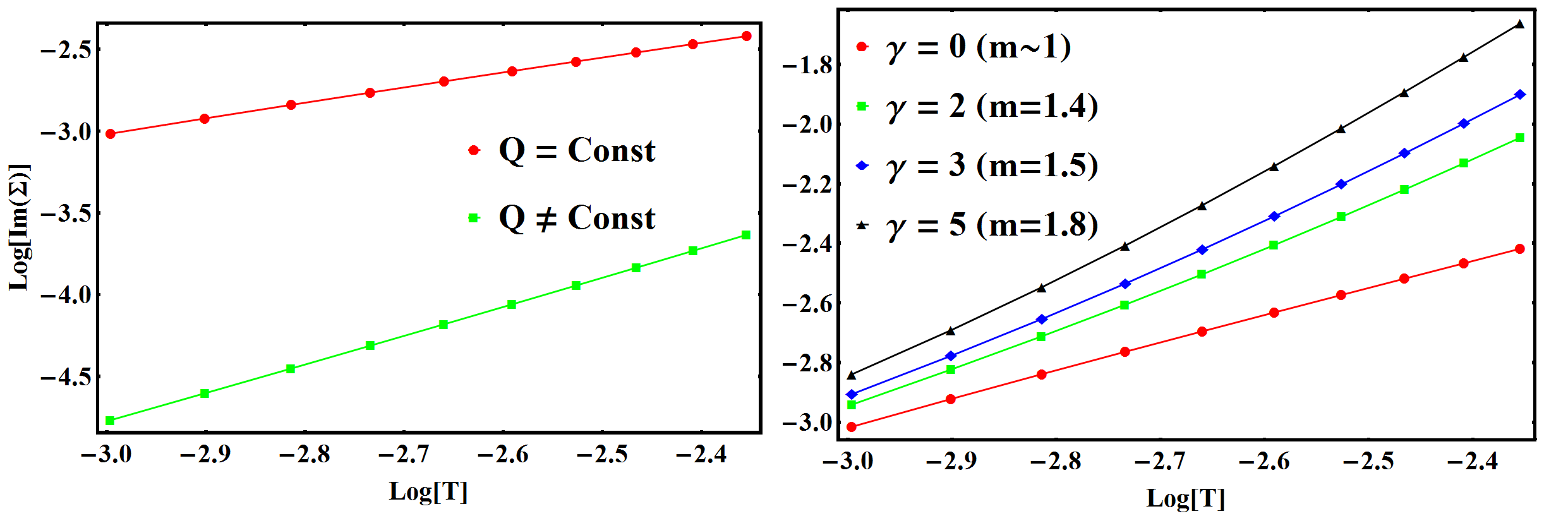}
\caption{A log-log plot of  imaginary part of the total self-energy versus temperature at the chemical potential $\mu=-0.273$ (case of perfect nesting). (Left) Comparison of the self  energy for the effective model when the bare susceptibility is approximated as a constant at the value of the nesting wave vector $Q = (0,\pi)$ to the case where the susceptibility has full momentum dependence. (Right) The  imaginary part of the total self-energy versus temperature for different values of the ratio $\gamma$ which is the ratio of the strength of intra-orbital to inter-orbital scattering. The quantity shown in the bracket is the slope ($m$) representing the exponent $n$. The value of $\omega$ is fixed at 2 meV.  }\label{ConstQ}
\end{figure}

 First, we note that the calculation in ref\cite{Ruvalds1990,Ruvalds1999} approximates the momentum dependent susceptibility to be a constant at the nesting wave vector $\vec Q$ throughout the Brillouin zone, presumably to maintain analytical tractability. A consequence of this approximation can be seen graphically in Fig \ref{Chi} where we plot the imaginary part of the total susceptibility in the Brillouin zone close to the nesting wave vector $\vec Q = (0, \pi)$. The plot on the right of Fig \ref{Chi} corresponds to the case of imperfect nesting where the imaginary part of the susceptibility lacks a peak at the nesting wave vector $\vec Q$. However, when nesting improves, the peak at the nesting wave vector is enhanced (Fig \ref{Chi} (left)). Therefore we conclude that the nesting approximation, which assumes the susceptibility to be a constant at the nesting wave vector under the perfect nesting condition ( i.e $\chi_0(\vec q, \omega) \approx \chi_0(\vec Q, \omega)$), highly overestimates the Coulomb scattering rate, and weakens the proposed $\omega/T$ scaling behavior. Under these conditions, it is already reasonable to expect a value of $n$ to be strictly greater than unity. This is confirmed in our numerical calculations of the Coulomb self energy in Fig \ref{ConstQ} (left) for the above two-band model. Clearly, there is an increase in the exponent $n$ to about 1.7 when the full momentum dependence of the susceptibility is inserted back into the calculation.
 At this point, it is important to understand that there is very little model dependence that can creep into this conclusion; the $(0, \pi)$ peak in the susceptibility under consideration is believed to be observed in most experiments and numerical calculations of higher orbital models in Iron superconductors (including $LiFeAs$), even in the absence of perfect nesting. As a result, there is hardly any reason to doubt that this approximation fails when the Fermi surfaces are perfectly nested. 

Second, in a multi-orbital system such as the pnictides, the contribution to the susceptibility in Eq.\ref{Susceptibility} comes from both intra-band and inter-band terms and these contributions cannot be disentangled. In the two-orbital model we considered, perfect nesting occurs only between two distinct bands$-$ the hole pocket at the $\Gamma$ point and the electron pocket at the $X$ point. Consequently, it is only these terms which contribute to non-Fermi liquid behavior, while the intra-band terms give the usual Fermi liquid physics. In an actual material, the ratio of the intra- and inter- band scattering strength determines the dominant behavior. This is shown in fig \ref{ConstQ}(right) which plots the imaginary part of the self energy as a function of temperature for different values of $\gamma = U_{intra}/U_{inter}$, where $U_{intra}, U_{inter}$ are the intra- and inter-orbital scattering strengths. With increasing values of $\gamma$, the curves become closer to that of a normal Fermi-liquid and is characterized by a slope which steadily moves away from unity. The fact that such intra-orbital terms exist in all Iron superconductors $-$ independent of the number of orbitals taken under consideration $-$ again, means that these conclusions hold in all models of $LiFeAs$.\\
\begin{figure}[h!]
\includegraphics[width=0.5\textwidth]{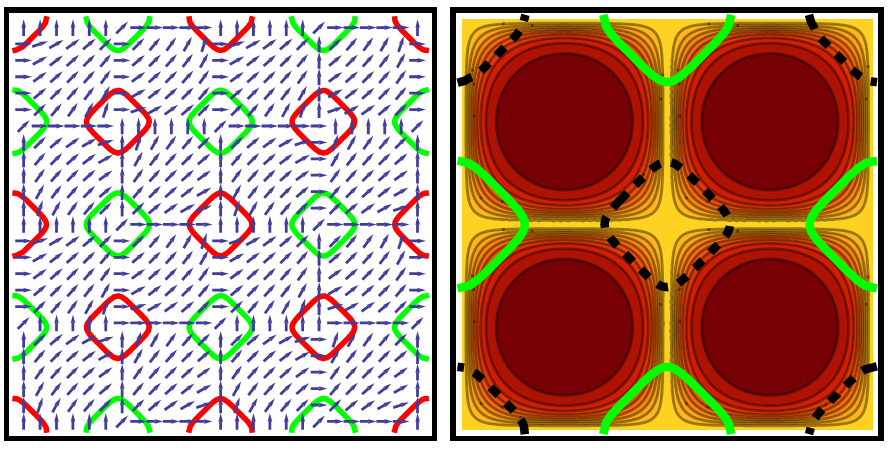}
\caption{Effect of orbital matrix elements on the efficiency of nesting. (Left) Orientation of the band angle $\theta_k$ throughout the extended Brillouin zone. Nesting between two disconnected pieces of the Fermi surface is efficient when their band angles differ by a value close to $\pi/2$ and minimal when they are in the same direction. This is more clearly seen in the right panel where the nesting probability is plotted in a color scale. The bright (dark) color represents regions where the probability is close to unity (zero). Therefore regions of the Fermi surface close to the $(\pi,0)$ and $(0,\pi)$ regions are efficiently nested. This must be contrasted with the single band case where the nesting probability is unity throughout the Brillouin zone.  }\label{MatrixElement}
\end{figure}
\newline
Third, it is evident from Eq.\ref{Susceptibility} that in a multi-orbital system, the susceptibility is weighted by the orbital matrix elements $\mathscr{L}_{mm'}^{\alpha \beta}(\vec k , \vec q)$. This is unlike a single orbital model where the weight factor is unity. To study their effect, we evaluate the matrix elements which are given by
\begin{equation}
\mathscr{L}_{mm'}^{\alpha \beta}(\vec k , \vec q) = U_{m\alpha}^*(\vec k + \vec q)  U_{m\beta}(\vec k + \vec q) U_{m'\beta}^*(\vec k) U_{m'\alpha}(\vec k),
\end{equation}
where $\hat{U}$ is the matrix which diagonalizes the band Hamiltonian given in Eq.\ref{BandHamiltonian}, $m,m'$ are band indices and $\alpha,\beta$ are orbital indices. As we are interested in the perfectly nested case, we want $m$ and $m'$ to be distinct corresponding to the hole pocket and electron pockets, or vice versa. In such a scenario, the matrix elements summed over the orbital indices are given by $\sum_{\alpha,\beta}\mathscr{L}_{mm'}^{\alpha \beta}(\vec k , \vec q) = sin^2(\theta_{\vec k+ \vec q} - \theta_{\vec k})$, with $m\neq m'$, and the band angle $\theta_k$  defined as
\begin{equation}
tan 2\theta_{k} = \frac{2 \epsilon_{xy}(\vec k)}{\epsilon_y(\vec k) - \epsilon_x(\vec k) }.
\end{equation}
 We can interpret the matrix element product as a nesting 'probability' because it takes on a positive value between zero and unity. Fig \ref{MatrixElement}(left) shows a vector plot of the band angle $\theta_k$ in the Brillouin zone defined with respect to the horizontal axis. As only the differences in band angles matter, we have plotted the angles shifted by $\pi/4$ to help visibility.
\begin{figure}[h!]
\includegraphics[width=0.5\textwidth]{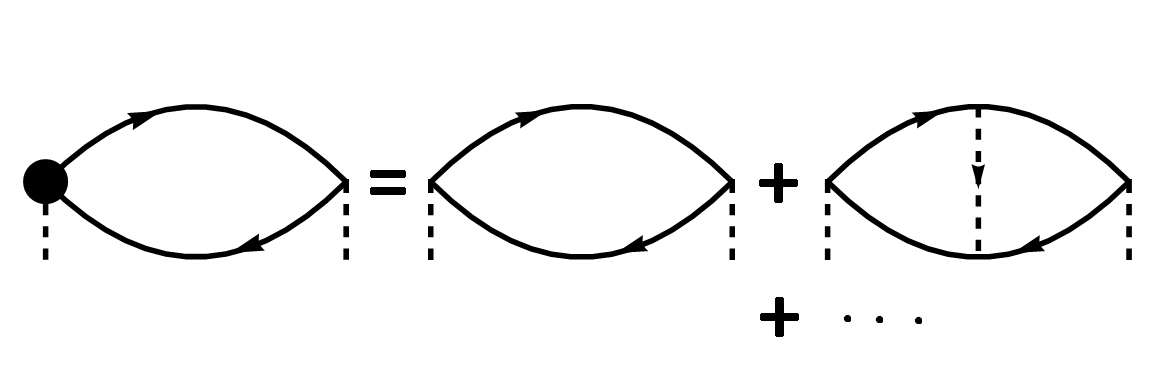}
\caption{Feynman diagrams showing the lowest order vertex corrections to the bare bubble susceptibility. The blob on the left represents the full renormalized vertex and all the fermion lines are treated as non-interacting. The dashed lines are the Coulomb interaction.}\label{VertexCorrection}
\end{figure}
From the above expression for the inter-orbital matrix element product, we conclude that in the above two orbital effective model, only those nested regions of the Fermi surface whose band angles differ by a value equal to $\pi/2$ contribute to nesting with maximum probability. This argument can be seen in a more straight forward manner in Fig \ref{MatrixElement}(right) which shows a color plot of the nesting probability in the Brillouin zone. Although at $\mu=-0.273$ both the hole and electron pockets are perfectly nested, regions on the Fermi surface along the diagonals contribute \textit{less} to the non-Fermi liquid-like behavior than regions along the $k_x$ and $k_y$ axis. This is very unlike the single band case where the nesting probability is equal to 1 for all regions of the nested Fermi surfaces. In a more accurate multiorbital description of $LiFeAs$, the matrix element product $-$ albeit has a very different form  $-$ still maintains a modulus value between zero and 1, and hence preserves the spirit of our conclusion. We have, therefore, shown that all the three factors discussed in the preceeding paragraphs increase the exponent $n$ to a value greater than unity\textit{ in a model independent manner}.

\begin{figure}[h!]
\includegraphics[width=0.5\textwidth]{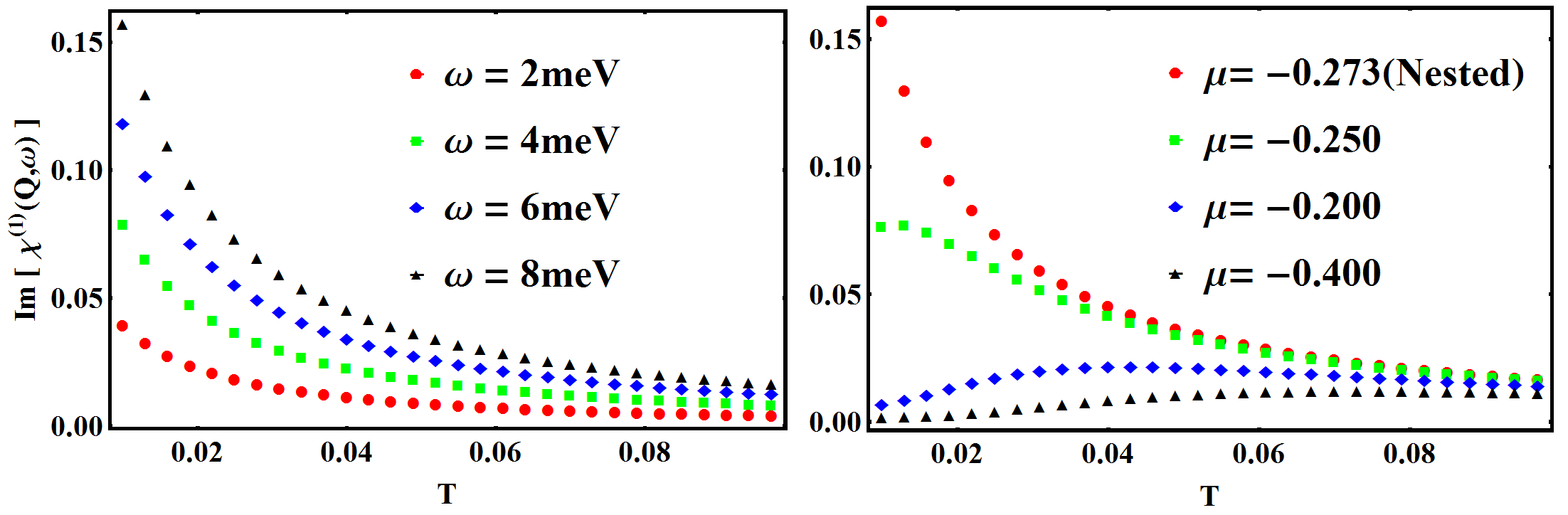}
\caption{Imaginary part of the lowest order vertex correction as a function of temperature: (left) plotted for the perfectly nested case ($\mu = -0.273$ eV) as a function of temperature for different values of $\omega$. A multi-orbital log divergence appears at small temperatures because intra-band effects are negligible. (Right) plotted for $\omega = 8$meV for different chemical potentials. The multi-orbital log divergence disappears as we move away from perfect nesting. }\label{Vertex}
\end{figure}

\textit{Beyond RPA- Vertex corrections:} In this section we study the effect of the lowest order vertex correction to the RPA self energy in the presence of nesting. Fig.\ref{VertexCorrection} shows the total vertex corrected diagram as a sum of the bare susceptibility along with the lowest and other higher order corrections. The lowest order vertex correction to the susceptibility is given by
\begin{eqnarray*}
\chi^{(1)}(\vec q, \omega) &=& \frac{1}{\beta V} \sum_{\vec p, i p_n} G^{(0)}_{uv}(\vec p, i p_n) G^{(0)}_{ab}(\vec p + \vec q, i p_n + i q_n)\\
&&\times \Gamma_{mnrs}(\vec p, \vec q, i p_n, i q_n),
\end{eqnarray*} 
where, $\Gamma_{mnrs}(\vec p, \vec q, i p_n, i q_n)$ is the orbital dependent correction to the vertex up to first order in the interaction, $\beta$ is the inverse temperature, $(a,b,u,v,m,n,r,s)$ are orbital indices, and the rest of the symbols have been previously defined. The lowest-order vertex correction is given by
\begin{eqnarray}
\Gamma_{mnrs}(\vec p, \vec q, i p_n, i q_n) &=& \frac{1}{\beta V} \sum_{\vec q', i q_n} (-V(\vec q'))\\\nonumber
&& \times G^{(0)}_{mn}(\vec p - \vec q' + \vec q, i p_n -  i q_n' + i q_n)\\ \nonumber
&&\times G^{(0)}_{rs}(\vec p - \vec q' , i p_n -  i q_n' ).\\ \nonumber
\end{eqnarray}
Substituting for the non-interaction Green functions from Eq.\ref{Green}, we obtain
\begin{widetext}
\begin{eqnarray}
\Gamma_{mnrs}(\vec p, \vec q, i p_n, i q_n) &=& \frac{1}{\beta V} \sum_{\substack{\vec q', i q_n'\\ \alpha, \beta}} (-V(\vec q'))  \frac{\mathscr{M}_{\alpha \beta}^{mnrs}(\vec p-\vec q', \vec q) }{\left(i p_n -  i q_n' +  i q_n - \mathscr{E}_{\alpha}(\vec p - \vec q' + \vec q) \right) \left(i p_n -  i q_n'  - \mathscr{E}_{\beta}(\vec p - \vec q' )\right )}\\ \nonumber
&=& \frac{1}{V} \sum_{\substack{\vec q'\\ \alpha, \beta}} V(\vec q') \mathscr{M}_{\alpha\beta}^{mnrs}(\vec p-\vec q', \vec q)  \left(  \frac{n_f\left(\mathscr{E}_{\beta}(\vec p -\vec q') \right) - n_f\left(\mathscr{E}_{\alpha}(\vec p - \vec q' + \vec q) \right)}{\mathscr{E}_{\alpha}(\vec p - \vec q' + \vec q) - \mathscr{E}_{\beta}(\vec p -\vec q') - i q_n }\right),
\end{eqnarray}
with the Matsubara sum over $i q_n'$ performed in the final step of the simplification. Substituting $\Gamma_{mnrs}(\vec p, \vec q, i p_n, i q_n)$ back into the previous expression and performing the remaining Matsubara sum over $ip_n$, we obtain the final expression of the lowest order vertex correction to the multi-orbital susceptibility:
\begin{eqnarray*}
\chi^{(1)}(\vec q, \omega) &=& \frac{-1}{V^2} \sum_{\substack{ \vec p, \vec q' \\ \alpha,\beta\\ \gamma, \delta}} \mathscr{M}_{\alpha\beta}^{mnrs}(\vec p-\vec q', \vec q) \mathscr{M}_{\gamma\delta}^{uvab}(\vec p, \vec q)  \left(  \frac{n_f\left(\mathscr{E}_{\beta}(\vec p -\vec q') \right) - n_f\left(\mathscr{E}_{\alpha}(\vec p - \vec q' + \vec q) \right)}{ \mathscr{E}_{\beta}(\vec p -\vec q')-\mathscr{E}_{\alpha}(\vec p - \vec q' + \vec q) +\omega + i\eta }\right)  \left(  \frac{n_f\left(\mathscr{E}_{\gamma}(\vec p) \right) - n_f\left(\mathscr{E}_{\delta}(\vec p + \vec q) \right)}{ \mathscr{E}_{\gamma}(\vec p )-\mathscr{E}_{\delta}(\vec p + \vec q) +\omega + i\eta }\right). \\ \nonumber
\end{eqnarray*}
\end{widetext}
We have defined the new matrix elements appearing above as
\begin{equation}
\mathscr{M}_{\gamma\delta}^{uvab}(\vec p, \vec q) = U_{\gamma u}(\vec p)^*  U_{\gamma v}(\vec p) U_{\delta a}(\vec p + \vec q)^* U_{\delta b}(\vec p + \vec q),
\end{equation}
with $(\alpha...\delta)$ as band indices, $\omega$ the probe frequency obtained after analytic continuation, and $\hat{U}$ the unitary orbital to band transformation matrix. Fig \ref{Vertex}(left) shows a numerical plot of the imaginary part of the vertex correction as a function of temperature for several $\omega$ for perfect inter-band nesting ($\mu = -0.273$).  All the curves show a logarithmic divergence at small values of $T$. As we will see in a moment, the divergence is entirely due to the inter-band nesting effect. Moreover, the effect of intra-band terms to the vertex correction is negligible due to the presence of this divergence. So one can safely neglect any intra-band terms unlike the zeroth order multi-orbital susceptibility seen before. Fig \ref{Vertex}(right) shows the same plot for different values of $\mu$ for fixed $\omega = 80$meV. As $\mu$ moves away from the perfectly nested value, the log divergence vanishes giving rise to a finite value at small values of $T$.\\
\newline
 To better understand this divergence at the nesting vector, we evaluate the expression for $\chi^{(1)}(\vec Q, \omega)$ for the simple case of a single orbital model. After converting the momentum sums into energy integrals, this is given by
\begin{eqnarray*}
\chi^{(1)}(\vec Q, \omega) &=& \frac{-N(0)^2}{(2\pi)^4}\int d\epsilon d\epsilon' g(\epsilon,\omega) g(\epsilon',\omega)\\
g(\epsilon,\omega) &=&  \left( \frac{tanh(\beta \frac{\epsilon}{2})}{2\epsilon + \omega + i \eta}\right),
\end{eqnarray*}
 with the range of the integrals from $-B/2$ to $B/2$ where $B$ is the bandwidth. Changing the integration variable to $x= \beta \epsilon/2$, defining $\nu = \beta \omega$, and taking the imaginary part we obtain
\begin{equation}
\chi^{(1)''}(\vec Q, \omega) = -4\pi N(0)^2 tanh\left(\frac{\nu}{4}\right) \int tanh(x) P\left( \frac{1}{4 x + \nu}\right) dx,
\end{equation}
where $P$ denotes the principal part and the limits of the integral are now from $-\beta B/4$ to $\beta B/4$. In the limit that the band width tends to infinity compared to $T$ and $T>>\omega$, we can write the above expression as
\begin{equation}
\chi^{(1)''}(\vec Q, \omega) \approx -4 \pi N(0)^2 tanh\left(\frac{\nu}{4}\right) Log\left(\beta B\right) + const,
\end{equation}
giving rise to the logarithmic divergence at $T\rightarrow0$ plus a constant independent of temperature.\\
\newline
To sum up, we have presented a study of how fractional powers of the resistivity can be understood from partial nesting between two disconnected pieces of the Fermi surfaces in multi-orbital systems. Through an effective two orbital model proposed for the Iron superconductors, we identified factors affecting the exponent $n$ in a \textit{model independent manner.} These factors included interband scattering, matrix element effects and invalidity of the commonly used nesting approximation which overestimates the scattering rate.  It is clear from our analysis of these factors that the exponent cannot reach the experimentally observed value of 1.35 even in the case of perfect nesting, calling for the presence of additional scattering mechanisms apart from a purely nesting driven mechanism. One could argue that such a conclusion is premature as a quantitative comparison could not be justified from just the study of an effective model such as that presented above. However, this discrepancy seems to be more serious in the hole doped $LiFeAs$ where the nesting between the inner hole pockets  at the $\Gamma$ point and those of the electrons pockets at the edge of the Brillouin zone gives rise to a transport exponent that is very close to unity\cite{Miao-private}. Given then, from our above calculations, that in a multiorbital system there are several factors pushing the value of $n$ to be greater than unity, such a statement would stand on a firm ground. Lastly, we calculated the role of lowest order vertex correction contribution to the susceptibility in the nested multi-orbital model and showed it to possess a logarithmic divergence that is strongly suppressed as the chemical doping weakens the nesting condition.\\
\newline
\textit{Acknowledgements:} CS  and PWP are supported by the Center for
Emergent Superconductivity, a DOE Energy Frontier Research Center,
Grant No. DE-AC0298CH1088.  PWP is also supported by the J. S. Guggenheim Foundation. We thank Garrett Vanacore for discussions and Mridula Damodaran for computational advice.
\bibliographystyle{apsrev4-1}
\bibliography{References}
\end{document}